\documentstyle[preprint,aps,epsf,floats]{revtex}    
\def\bq{\begin{equation}}
\def\eq{\end{equation}}
\def\ba{\begin{eqnarray}}
\def\ea{\end{eqnarray}}
\begin{document}
\thispagestyle{empty}
  
\renewcommand{\small}{\normalsize} 

\preprint{
\font\fortssbx=cmssbx10 scaled \magstep2
\hbox to \hsize{
\hskip.5in \raise.1in\hbox{\fortssbx University of Wisconsin - Madison} 
\hfill\vtop{\hbox{\bf MADPH-99-1143}
            \hbox{November 1999}} }
}
\title{\vspace{.5in}
$b$-quark decay in the collinear approximation
}

\author{D.~Zeppenfeld}
\address{
Department of Physics, University of Wisconsin, Madison, WI 53706 
}
\maketitle
\begin{abstract}
The semileptonic decay of a $b$-quark, $b\to c\ell\nu$, is considered in the 
relativistic limit where the decay products are approximately collinear.
Analytic results for the double differential lepton energy distributions 
are given for finite charm-quark mass. Their use for the fast simulation 
of isolated lepton backgrounds from heavy quark decays is discussed.
\end{abstract}
%
%

\newpage

%
%

Many new physics signals at hadron colliders involve isolated hard leptons
as a distinguishing feature. Sequential decays of supersymmetric 
particles~\cite{susy.lepton} and Higgs boson decays to $Z$, $W$ or $\tau$
pairs~\cite{CMS-ATLAS,Hww} are but two examples. In all these cases 
the production of heavy quarks, in particular bottom and charm, and their 
subsequent semileptonic decay constitutes
an important background. Even though lepton isolation, the requirement that
little hadronic energy is deposited in the vicinity of the charged decay 
lepton, can reduce these heavy quark backgrounds by large factors, the sheer
size of the $b\bar b$ or $c\bar c$ production cross section makes heavy
flavor backgrounds dangerous~\cite{CMS-ATLAS}.

For the simulation of such heavy flavor backgrounds the large suppression 
factors due to lepton isolation pose a special problem: large Monte Carlo 
samples must be generated in order to analyze the phase space distributions 
of the surviving events. While the full five-dimensional distribution of 
$b\to c\ell\nu$ decay is easily implemented in a Monte Carlo 
program~\cite{earlyHQ}, this procedure does not always generate isolated 
lepton events in a sufficiently fast and efficient manner.

In this brief note I describe how a fast short-cut is provided by 
analytic expressions for the lepton energy
distributions in the laboratory frame. Leptons of sufficiently high transverse 
momentum can only result from the decay of very energetic $b$ or $c$ quarks. 
In turn, this implies that the parent quarks must be moving relativistically
in the lab, which results in the decay products moving approximately collinear
to the parent quark direction. In this relativistic limit, only 
the energy fractions of the decay particles, as compared to the heavy quark
energy, are needed for a full description.

To be definite, consider the decay $b\to c\ell\nu$ and denote the energy
fractions of the neutrino, the charged lepton and the $c$-quark by
\bq
x = {E_\nu\over E_b}\;, \qquad y = {E_\ell\over E_b}\;, \qquad
z = {E_c \over E_b}\;, 
\eq
respectively. Obviously they obey the constraint $x+y+z=1$. The smallest
energy for the charm quark is reached when, in the $b$ rest frame, it is 
emitted opposite to the 
$b$-quark direction, recoiling against a collinear lepton-neutrino pair:
\bq
z \geq r = {m_c^2\over m_b^2} \; .
\eq
In the spectator quark model, and for unpolarized $b$-quarks,
the double differential $b$-decay distribution can be determined
analytically. I find
\bq
\label{eq:d2Gdxdy}
{1\over \Gamma} {d^2\Gamma\over dx dy} = {2c\over f(r)}\;
\biggl( c\;(1-x)\;\left[ c+(3-c)\;x\; \right] +
3ry\;{(2-c)\;x+c \over 1-x-y}\; \biggr) \; .
\eq
Here 
\bq
c = {1-r-x-y\over 1-x-y} = 1-{r\over z} \;,
\eq
and $f(r)$ is the phase space suppression factor for the $b\to c\ell\nu$
decay due to the finite charm quark mass~\cite{fofr,tsai},
\bq
\label{eq:width}
f(r) = (1-r^2)(1-8r+r^2) - 12r^2\log r \; ,
\eq
which is quite sizable for $b$-decay: $f(r)=0.42$ for $r=0.12$.

For $c\to s\ell\nu$ decay the $(V-A)\times (V-A)$ structure of the weak decay 
amplitude implies a double differential decay distribution identical to
Eq.~(\ref{eq:d2Gdxdy}), but with the role of charged lepton and neutrino energy
fractions interchanged, i.e. for charm decay $x=E_\ell/E_c$, $y=E_\nu/E_c$,
$z=E_s/E_c\geq r=m_s^2/m_c^2$. 

In the massless limit, $m_c= 0$ i.e. $r=0$ and $c=1$, the double differential 
distribution of Eq.~(\ref{eq:d2Gdxdy}) reduces to 
\bq
\label{eq:d2Gdxdym0}
{1\over \Gamma_0} {d^2\Gamma_0\over dx dy} = 
2\; (1-x)\;\left( 1+2x \right) \; ,
\eq
which leads e.g. to the well known lepton decay distribution in 
$\tau \to\ell\bar\nu_\ell\nu_\tau$ decay~\cite{tsai,tau}
\bq
\label{eq:dGdym0}
{1\over \Gamma_0} {d\Gamma_0\over dy} = 
\int_0^{1-y}dx\; {1\over \Gamma_0} {d^2\Gamma_0\over dx dy} = 
{1\over 3}
(1-y)\;\left( 5+5y-4y^2 \right) \; .
\eq

The double differential decay distribution of Eq.~(\ref{eq:d2Gdxdy}) provides
an adequate description of charged lepton and missing transverse momentum 
distributions and their correlations in typical collider physics 
applications. Because of its simple algebraic form, 
it may be folded analytically with algebraic fragmentation functions, like the 
Peterson fragmentation function~\cite{peterson} for $b$-quarks. 
Also, the collinear limit is sufficiently simple to cast phase space limits
into limits on the momentum fractions of the $b$-decay
products, once the momentum of the parent $b$ is known, e.g. in a 
Monte Carlo program. 

Another application 
is the effect of lepton isolation on the observable 
charged lepton or missing transverse momentum distributions in $b$ decays. 
A typical lepton isolation cut limits the energy fraction carried by the 
charm quark, to e.g. 10~\% of the observable lepton energy, or imposes an 
upper limit, e.g. 5~GeV, on its transverse energy. Taking $m_b = 5.28$~GeV 
and $m_c=1.87$~GeV, i.e. using the lightest meson masses in order to 
approximately obtain the correct kinematics for the heavy quark decays,
one finds $z>r=12.5\%$, which, at face value, excludes any events where the
charm quark would carry as little as 10\% of the charged lepton energy
or would limit the $b$-quark $E_T$, and thereby the maximum lepton $E_T$
to about 40~GeV when $E_{Tc}<5$~GeV is required.
However, these limits 
are imposed in the experiment on observed hadrons, or calorimeter response 
in some cone around the lepton direction. For the soft hadronic depositions 
inside the lepton isolation cone, non-perturbative corrections 
(from fragmentation or underlying event contributions) or fluctuations in 
the calorimeter response lead to considerable uncertainties in the true 
energy fraction $z$ carried by the charm quark. The low energy
tails of the calorimeter response to charm quarks are largely responsible
for fake isolated lepton events. As a result, the actual $z$-distribution of
the charm quark requires detailed simulations, except for the general 
statement that small values of $z$, close to their
kinematic limit $z=r$, are strongly favored. 

The double differential decay distribution derived above allows to assess
the effects that $z$-smearing has on the observed lepton distributions. At 
fixed $z=1-x-y$ we may study the charged lepton energy distribution
\bq
{1\over \Gamma_z}{d\Gamma_z\over dy}(y) = {1\over N(z)}
{1\over \Gamma} {d^2\Gamma\over dx dy}(x=1-y-z,\; y) \;,
\eq
or the analogous neutrino energy distribution 
$1/ \Gamma_z\;d\Gamma_z/ dx$.
Here $N(z)$ is a normalization factor which is obtained by direct integration
of Eq.~(\ref{eq:d2Gdxdy}):
\ba
\label{eq:dGdz}
N(z) &=& {1\over \Gamma} {d\Gamma\over dz} = 
\int_0^{1-z}dx \; {1\over \Gamma} {d^2\Gamma\over dx dy}(x,y=1-x-z)  
\nonumber \\
&=& {2\over f(r)}\;(1-{r\over z})
(1-z)\biggl( (1-r)^2 + {1\over 6}\;
(1-{r\over z})(1-z)\;(4z-r-1+4{r\over z})\biggr)\; .
\ea
\begin{figure}[t]
\vspace*{0.5in}            
\begin{picture}(0,0)(0,0)
\includegraphics{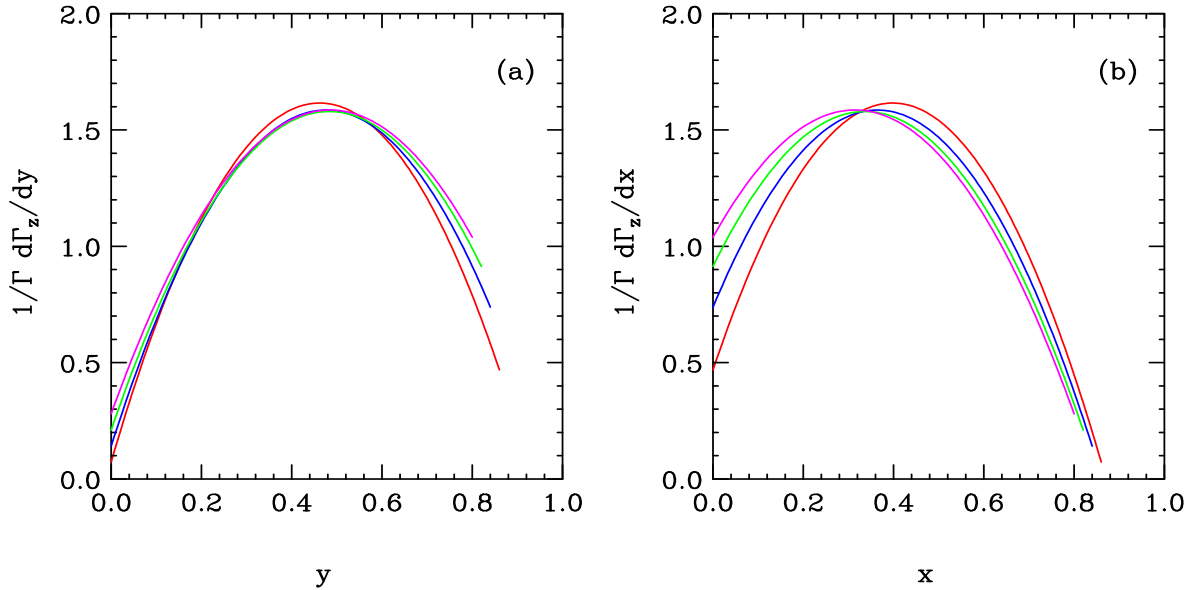}
\end{picture}
\vspace{7.5cm}
\caption{Normalized energy distributions of (a) the charged lepton and 
(b) the neutrino in $b\protect\to \nu\ell c$ decays in the spectator model. 
The individual curves correspond to fixed charm quark energy fractions 
$z = 0.14$ (red), 0.16 (blue), 0.18 (green), 0.2 (magenta), 
which are slightly above threshold, given by $z=r=m_c^2/m_b^2=0.12$.
}
\vspace*{0.2in}
\label{fig1}
\end{figure}

These neutrino and charged lepton energy distributions are shown in 
Fig.~\ref{fig1}. They change very little with $z$, in the $z$-range
leading to isolated leptons. The largest $z$-dependence is found near 
the kinematic limits, somewhat affecting the hardest charged leptons and
the softest neutrinos. The modest $z$-dependence of the lepton distributions
implies that the precise $z$-distribution produced by the lepton isolation 
cuts is not needed for an adequate description of lepton momentum 
distributions.

One thus finds that in $b$-quark decays which lead to isolated leptons, 
the charged lepton and missing transverse momentum distributions, and their 
correlations, can be 
modeled quite reliably, making use of the double differential decay
distribution described here. A first application in collider phenomenology
appears in Ref.~\cite{prz} where $b\bar b+$~jets backgrounds to 
$H\to\tau\tau$ searches are discussed: the decay distributions
in the collinear approximation allow to considerably
improve Monte Carlo statistics. Similar improvements are foreseen in 
the simulation of heavy quark backgrounds for many new physics signals 
involving isolated charged leptons.

%

\acknowledgements
I would like to thank M.~G.~Olsson for discussions which greatly helped
deriving Eq.~(\ref{eq:d2Gdxdy}).
This research was supported in part by the University of Wisconsin 
Research Committee with funds granted by the Wisconsin Alumni Research 
Foundation and in part by the U.~S.~Department of Energy under 
Contract No.~DE-FG02-95ER40896.

%
%

\end{document}